\documentclass[aps,reprint,showpacs,showkeys,prl,superscriptaddress]{revtex4-2}
\usepackage{dcolumn}
\usepackage[utf8]{inputenc}
\usepackage[english]{babel}
\usepackage{csquotes}
\usepackage{amsmath}
\usepackage{amsthm}
\usepackage{amsfonts}
\usepackage{amssymb}
\usepackage{graphicx}
\usepackage{soul}
\usepackage[skip=0.5ex]{subcaption}
\usepackage{hyperref}
\usepackage{ulem,xcolor,tikz}
\usepackage{braket}
\usepackage{mathtools}
\usepackage{bbm}
\usepackage[toc]{appendix}
\usepackage{verbatim}

\usepackage{ragged2e}

\newcommand{\dd}{\mathrm{d}}

\allowdisplaybreaks 

\begin{document}

\title{Surpassing the wave-particle duality relation via feed-forward of phase information}

\newcommand*{\QOQI}{Quantum Optics and Quantum Information Group, \\Friedrich-Alexander-Universität Erlangen-Nürnberg, Staudtstr.~1, 91058 Erlangen, Germany}

\author{Elisabeth Meusert}
\thanks{These two authors contributed equally.}
\author{Uwe Schilling}
\thanks{These two authors contributed equally.}
\author{Marc-Oliver Pleinert}
\author{Joachim von Zanthier}
\affiliation{\QOQI}

\begin{abstract}

Complementarity constitutes a central aspect of quantum theory. It manifests itself, for example, in a two-way interferometer, where the simultaneous observation of an interference pattern and the acquisition of which-way information are limited by an inequality known as the duality relation.
Here, we investigate which-way information in a double-slit interferometer, which can be correlated to the phase of the quantum object at the detection screen, leading to a phase-dependent which-way knowledge. In specific cases, this knowledge can locally exceed the limit set by the duality relation. Based on this observation, we propose a feed-forward protocol that aims at maximizing the which-way information locally for each phase after the particle has been recorded on the screen. This allows us to surpass the duality relation limit even globally.
We present analytical results as a proof of principle of our protocol as well as numerical outcomes quantifying the amount of maximally achievable which-way knowledge.

\end{abstract}

\maketitle

The concept of \textit{complementarity} lies at the heart of quantum theory and has been put forward already by Einstein and Bohr at its early stages \cite{Bohr:1949}.
It states that quantum systems possess pairs of properties, i.e., conjugate variables,
which are equally real but mutually exclusive.
Their mutual exclusivity manifests itself in the non-commutability of their corresponding observables, i.e., the impossibility of observing both complementary properties simultaneously.

A well-known example for two complementary observables occurs for a quantum object (QO) passing through a two-way interferometer.
Here, the simultaneous observation of an interference pattern and the acquisition of full which-way (WW) information of the QO is excluded. The complementary observables are associated with wave- and particle-like behaviour and quantified through the \textit{visibility} $\mathcal{V}$ of the resulting interference pattern and the \textit{which-way knowledge} $\mathcal{K}_{\hat{W}}$, respectively.
The subscript $\hat{W}$ of the WW knowledge indicates its dependence  on the choice of the observable for the which-way detector, which we will derive later.
As has been shown independently by Englert \cite{Englert:1996} and Jaeger \textit{et al.}~\cite{Jaeger:1995}, the complementarity of the two quantities can be quantified via the \textit{duality relation}
\begin{align}
\mathcal{K}_{\hat{W}}^2 + \mathcal{V}^2 \leq 1.
\label{eq_duality}
\end{align}
This inequality has been confirmed in numerous experiments~\cite{Durr:1998,Durr:1998a,Buks:1998,Schwindt:1999,Pryde:2004,Peng:2005,Jacques:2008,Barbieri:2009} and theoretical investigations~\cite{Bjork:1998,Abranyos:1999,Englert:2000,Miniatura:2007,Erez:2009}, and various aspects of it are subject to ongoing research, e.g., its validity in the presence of a coupling to an environment~\cite{Bolduc:2014, Leach:2016} or in asymmetric interferometers~\cite{Chen:2022}, among others~\cite{Schilling:2012,Bera:2015,Spegel-Lexne:2024,Jiang:2025}

The duality relation \eqref{eq_duality} implies that any availability of WW knowledge about the quantum object's path causes a reduction of the interference pattern's visibility, i.e., the quantum object's ability to interfere with itself.
Vice versa, given the visibility of the interference pattern, the amount of WW knowledge that is \textit{principally available}, commonly termed \textit{distinguishability}, is upper bounded by $\sqrt{1-\mathcal{V}^2}$.
The problem of how to achieve this optimal WW knowledge has been solved~\cite{Englert:1996}.
 
Here, we present a protocol that enables us to exceed this limit via feed-forward of phase information.
To do so, we derive a dependence of the WW knowledge on the observable chosen for the readout at the which-way detector (WWD).
We then observe that, in general, the amount of WW knowledge obtained is correlated to the QO's phase~\cite{Schilling:2012} and can, in certain cases, lead to a local surpassing of the duality relation.
This enables us to maximize the WW knowledge individually for each experimental run by first detecting the QO after passage through the interferometer at the screen.
The resulting phase information is then feed-forwarded and utilized to choose the WWD observable that returns the highest possible WW knowledge for the found phase.
Finally, we demonstrate that this protocol allows to surpass the limit set by the duality relation \eqref{eq_duality}. We  present analytical and numerical results for the increase of WW knowledge achieved compared to the previously known optimization strategy~\cite{Englert:1996}.

\paragraph*{Interferometer setup.}

\begin{figure*}
\centering\includegraphics[width=\textwidth]{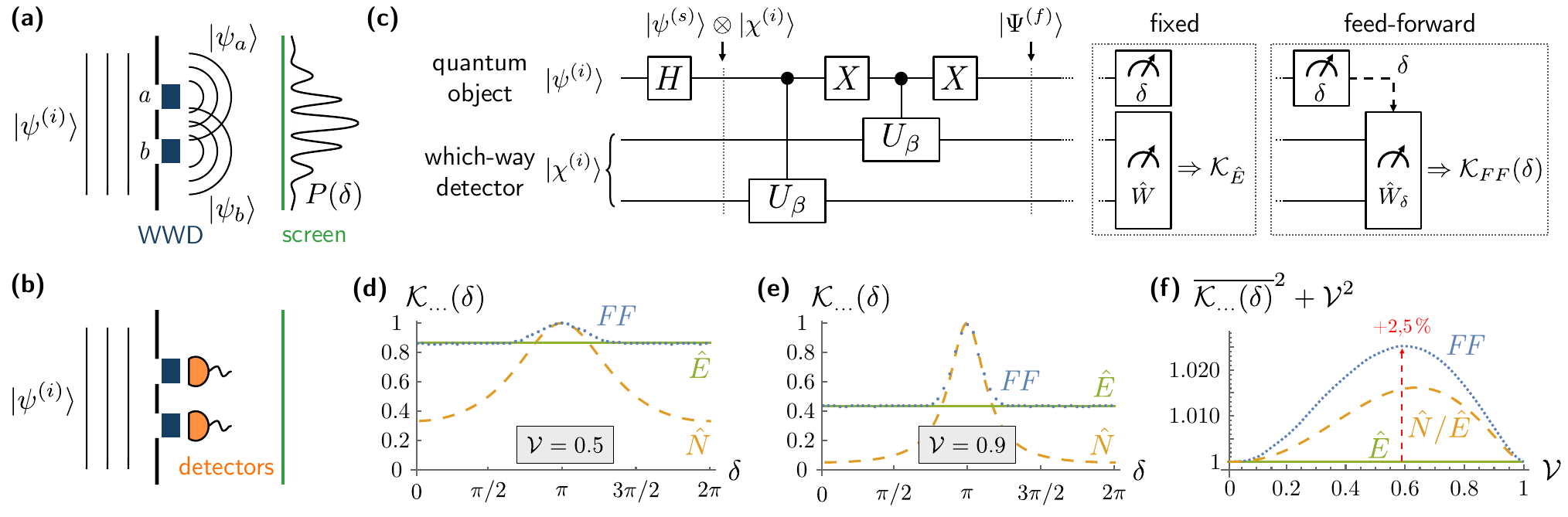}
\caption{\justifying (a) Young-type double slit interferometer consisting of two pathways $a$ and $b$, two which-way detectors (WWD) within each pathway and a screen with an interference pattern $P(\delta)$.
(b) Setup for experimental verification of the which-way knowledge.
(c) Comparison of a fixed basis choice and the presented feed-forward protocol in a quantum circuit diagram. 
(d,e) 
Phase-dependent knowledge $\mathcal{K}_{\ldots}(\delta)$ (canonical $\hat{E}$: solid; natural $\hat{N}$: dashed; feed-forward $FF$: dots) for two different visibilities (d) $\mathcal{V} = 0.5$ and (e) $\mathcal{V} = 0.9$.
(f) Sum of the phase-averaged knowledge $\overline{\mathcal{K}_{\ldots}(\delta)}$ squared and visibility $\mathcal{V}$ squared for the canonical knowledge ($\hat{E}$, solid), the simplified version of the feed-forward protocol ($\hat{N}/\hat{E}$, dashed), and the non-simplified feed-forward protocol ($FF$, dotted). 
}
\label{fig_all}
\end{figure*}

Throughout this paper, we consider a Young-type double slit interferometer as shown in Fig.~\ref{fig_all}(a) for conceptual simplicity; however, the formalism is applicable to any type of two-way interferometers, e.g., a Mach-Zehnder interferometer.
In such a setup, a QO enters the interferometer with initial wavefunction $|\psi^{(i)}\rangle$. At the double slit, it is equally split into 
\begin{align}
    |\psi^{\left(i\right)}\rangle \xrightarrow[\text{passage}]{\text{slit}} \ket{\psi_a}+\ket{\psi_b} \equiv  \ket{\psi^{(s)}} \, ,
    \label{eq_QO_wavefunc_slit}
\end{align}
where the two orthogonal states $\ket{\psi_a}$ and $\ket{\psi_b}$ are associated with the two pathways $a$ and $b$ of the QO through the interferometer. 
Note that we omit normalization factors in the states throughout this paper.
Upon propagation towards a position $\mathbf{r}$ on the detection screen, the parts $\ket{\psi_a}$ and $\ket{\psi_b}$ accumulate phases $\phi_a(\mathbf{r})$ and $\phi_b(\mathbf{r})$, resulting in a relative phase between the two paths $\delta \equiv \delta(\mathbf{r}) = \phi_b(\mathbf{r}) - \phi_a(\mathbf{r})$.
The state of the QO at position $\mathbf{r}$ with corresponding phase $\delta$ is thus given by $\ket{\psi_\delta}\equiv \ket{\psi_a}+e^{i\delta}\ket{\psi_b}$.

\paragraph*{Which-way detector.}

To gain information about the path of the QO, we introduce a which-way detector (WWD) at the interferometer slits, depicted as two boxes in the two pathways of Fig.~\ref{fig_all}(a). We assume that each part of the WWD is implemented by a qubit initialized in the state $|\chi^{(i)}\rangle = \ket{0}_a\otimes\ket{0}_b \equiv \ket{00}$. Assuming that the WWD performs a quantum non-demolition measurement \cite{Scully:1991} of the QO's path,  the combined system after interaction of the QO with the WWD is transferred into the entangled state~\cite{Schilling:2012}:
\begin{align}
    \ket{\psi^{(s)}} \otimes |\chi^{\left(i\right)}\rangle \overset{\text{WWD}}{\longrightarrow}\ 
     \ket{\psi_a}\ket{\chi_a}+\ket{\psi_b}\ket{\chi_b} \equiv |\Psi^{\left(f\right)}\rangle \, . 
    \label{eq_QO_WWD_entangledstate}
\end{align}
Hereby, the two final states $|\chi_a\rangle$ and $|\chi_b\rangle$ of the WWD are, in contrast to the states $|\psi_a \rangle$ and $|\psi_b \rangle$ of the QO, not necessarily orthogonal. Instead, they are a general superposition of the basis states $\ket{0}_{a/b}$ and $\ket{1}_{a/b}$:
\begin{align}
    \nonumber & \ket{\chi_a} = (\alpha\ket{0}_a+\beta\ket{1}_a)\otimes\ket{0}_b = \alpha\ket{00}+\beta\ket{10},\\
    & \ket{\chi_b} = \ket{0}_a\otimes (\alpha\ket{0}_b+\beta\ket{1}_b) = \alpha\ket{00}+\beta\ket{01}
    \label{eq_WWD_std-basis}
\end{align}
with $\alpha, \beta \in \mathbb{C}$ and $|\alpha|^2+|\beta|^2=1$~\footnote{In general, $\ket{\chi_{a,b}}$ may have a relative phase $e^{i\phi}$, which besides shifting the interference pattern by a constant amount $\phi$ is without any further effect.}. $\beta$ governs the interaction strength between the QO and the WWD \cite{Schilling:2012}, which we assume to be equal for both qubits. A value of $|\beta| < 1$ allows for a partial measurement of WW knowledge with a principally suboptimal WWD.

\paragraph*{Interference pattern and which-way knowledge.}

The state of the QO after interaction with the WWD is described by the density matrix $\rho_\text{QO}= \text{tr}_\text{WWD}(|\Psi^{(f)}\rangle\langle\Psi^{(f)}|)$, and the resulting interference pattern is given by
\begin{align}
    P(\delta) = \langle\psi_\delta|\rho_\text{QO}|\psi_\delta\rangle = \frac{1}{2\pi}\left[1+|\langle\chi_a|\chi_b\rangle|\,\cos{(\delta)}\right]
    \label{eq_QO_interference_WWD}
\end{align}
with the normalization chosen to be done over one $2\pi$-period.
Its visibility results in
$\mathcal V = |\langle \chi_a | \chi_b\rangle | = |\alpha|^2$.
Hence, the smaller the overlap of $|\chi_a\rangle$ and $|\chi_b\rangle$, the easier it is to distinguish between the two paths and the lower is the visibility. For example, if $|\chi_a\rangle$ and $|\chi_b\rangle$ are orthogonal, reading out the WWD in this basis~\footnote{Strictly speaking, the WWD Hilbert space has dimension $\text{dim}(\mathbb{C}^2\otimes\mathbb{C}^2) = 4$ and $\{\ket{\chi_a},\ket{\chi_b}\}$ alone cannot form a basis. 
However, in case of orthogonal $\ket{\chi_{a/b}}$, this set can be completed to an eigenbasis of a WWD observable that returns full WW information.} will give full information about the QO's path while no interference pattern will appear. 
However, since in general $\ket{\chi_a}$ and $\ket{\chi_b}$ are not orthogonal, they cannot be simultaneous eigenstates of any observable $\hat{W}$ of the WWD. We thus first have to specify an observable $\hat{W}$ and its eigenbasis $W = \{|W_i\rangle,\ i=1,\ldots,n\}$ in order to express $|\chi_a\rangle$ and $|\chi_b\rangle$ in terms of this eigenbasis. As a consequence, the readout of $\hat{W}$ provides, in most cases, incomplete information about the path of the QO, that can be quantified in the following way.

A measurement of the WWD observable $\hat{W}$ returns an eigenvalue $W_i$, leaving the WWD in eigenstate $\ket{W_i}$, with probability 
\begin{align}
    p_i = \langle W_i|\rho_\text{WWD}|W_i\rangle,
    \label{eq_pi}
\end{align}
where $\rho_\text{WWD} = \text{tr}_\text{QO}(|\Psi^{(f)}\rangle\langle\Psi^{(f)}|)$
is the WWD's state after passage of the QO. Assuming that such a measurement has been realized, the state of the QO-WWD system is projected onto $\ket{W_i}$, resulting in
\begin{align}
    |\Psi^{(f)}\rangle \xrightarrow[\text{readout}]{\text{WWD}} \langle W_i|\chi_a\rangle \ket{\psi_a} + \langle W_i|\chi_b\rangle \ket{\psi_b} \equiv \ket{\psi_i}. \notag
\end{align}
If we decided to subsequently check the QO's path with a setup as shown in Fig.~\ref{fig_all}(b), we would find the QO behind path $a$ or $b$ with probability $|\langle\psi_{a/b}|\psi_i\rangle|^2 \sim |\langle W_i| \chi_{a/b}\rangle|^2$.
The probabilities of guessing the path correctly by stating that the QO has taken path $a$ or $b$ after the WWD has been found in eigenstate $\ket{W_i}$ are thus proportional to $|\langle W_i|\chi_{a/b}\rangle|^2$.
We can therefore express the highest possible probability of guessing the path correctly from $\ket{W_i}$ as~\cite{Schilling:2012}
\begin{align}
    q_i = \frac{\text{max}\left(|\langle W_i|\chi_a\rangle|^2,\ |\langle W_i|\chi_b\rangle|^2\right)}{|\langle W_i|\chi_a \rangle|^2+|\langle W_i|\chi_b\rangle|^2}.
    \label{eq_qi}
\end{align}
This probability serves as a measure for the WW information obtained in a single run of the experiment. However, during the readout process, the WWD is reduced to one of its eigenstates $\ket{W_i} \in W$ randomly.
In order to obtain a measure for the 
\textit{likelihood}
$\mathcal{L}_{\hat{W}}$
of guessing the QO's path correctly in an arbitrary run of the experiment in which we measure the WWD observable $\hat{W}$, we need to average over all possible eigenstates $\ket{W_i}$ with their respective probabilities $p_i$, i.e., $\mathcal{L}_{\hat{W}} = \sum_i p_iq_i$.
Note that the value of $\mathcal{L}_{\hat{W}}$ runs from $0.5$ to $1$: If we don't know anything about the QO's path, we need to make a random guess, which is correct with a probability of $0.5$, while if we know the path for sure, we are correct with probability $1$.
To obtain a quantity that returns $0\ (1)$ in case of no (full) WW information, we rescale $\mathcal{L}_{\hat{W}}$ to the \textit{which-way knowledge}~\cite{Englert:1996,Schilling:2012}
\begin{align}
    \mathcal{K}_{\hat{W}} \equiv 2\mathcal{L}_{\hat{W}}-1 = 2\sum_i p_iq_i - 1\, , \label{eq_knowledge}
\end{align}
which, by construction, adheres to the duality relation given in Eq.~\eqref{eq_duality} \cite{Englert:1996}.
The subscript $\hat{W}$ highlights the dependency on the chosen WWD observable.

\paragraph*{Natural and canonical observable.}\label{sec:nat-and-can-observables}

As a straightforward example, we consider the \textit{natural observable} $\hat{N}$, where we read out each qubit locally, i.e., in the eigenbasis $N = \{\ket{00}, \ket{10}, \ket{01}, \ket{11}\}$. This basis results in the \textit{natural knowledge}
\begin{align}
    \mathcal{K}_{\hat{N}} = |\beta|^2 = 1-\mathcal{V} \leq \sqrt{1-\mathcal{V}^2},
    \label{eq_knowledge_natural}
\end{align}
which for $|\beta|^2 < 1$ ($\mathcal{V} > 0$), i.e., imperfect WWD sensitivity, is less than the optimal value allowed by the duality relation \eqref{eq_duality}. In order to maximize the WW knowledge obtained on average for a fixed WWD observable choice, one has to choose the \textit{canonical observable} $\hat{E} = |\chi_a\rangle\langle\chi_a|-|\chi_b\rangle\langle\chi_b|$ \cite{Englert:1996}. Its eigenvectors return the \textit{canonical knowledge}
\begin{align}
    \mathcal{K}_{\hat{E}} = \sqrt{1-|\alpha|^4} = \sqrt{1-\mathcal{V}^2},
    \label{eq_knowledge_canonical}
\end{align}
which is the optimal value permitted by the duality relation~\eqref{eq_duality}. The canonical knowledge is also often called the distinguishability $\mathcal{K}_{\hat{E}} \equiv \mathcal{D}$~\cite{Englert:1996}.

\paragraph*{Phase-dependent which-way knowledge.}

As stated in the introduction, the dual nature of WW knowledge and visibility is only displayed if both quantities are measured within the same set of experimental runs.
Under a \textit{single run of the experiment} we understand: the passing of a QO through the double slit, its interaction with the WWD, and the subsequent detection of the QO at the screen as well as the readout of the WWD observable $\hat{W}$.

After slit passage and interaction with the WWD, the QO-WWD system is in state $|\Psi^{(f)}\rangle$. The detection of the QO at the screen, i.e., the measurement of its phase $\delta$, projects that state onto $\ket{\psi_\delta}$:
\begin{align}
    |\Psi^{(f)}\rangle 
    \xrightarrow[\text{measurement}]{\text{phase}}
    \langle\psi_\delta|\Psi^{(f)}\rangle.
    \label{eq_QO_WWD_phase-measurement}
\end{align}
When subsequently measuring the WWD observable $\hat{W}$, this state is projected \textit{further} onto the resulting eigenstate. 
The conditional probability of finding a specific eigenstate $\ket{W_i}$ having found the QO at phase $\delta$ is then given by 
\begin{align}
p_i(\delta) = \frac{|\langle \psi_\delta,W_i|\Psi^{(f)}\rangle|^2}{P(\delta)}
    \label{eq_pi_delta}
\end{align}
with $|\psi_\delta,W_i\rangle= |\psi_\delta\rangle \otimes |W_i\rangle$  
and $P(\delta)$ from Eq.~\eqref{eq_QO_interference_WWD}.
Having found the WWD in an eigenstate $\ket{W_i}$, the probability of guessing the QO's path correctly is still given by $q_i$ from Eq.~\eqref{eq_qi}. 
However, the phase-dependency of $p_i(\delta)$ 
leads to a \textit{phase-dependent} WW knowledge \cite{Schilling:2012}:
\begin{align}
    \mathcal{K}_{\hat{W}}(\delta) = 2\sum_i p_i(\delta)q_i-1 .
    \label{eq_knowledge_phasedep}
\end{align}
$\mathcal{K}_{\hat{W}}(\delta)$ returns the knowledge obtained on average for experiments where the QO was found at a \textit{specific} $\delta$. 
In contrast, $\mathcal{K}_{\hat{W}}$ from Eq.~\eqref{eq_knowledge} returns the knowledge obtained on average for experiments where the QO was found at an \textit{arbitrary} $\delta$. 
Thus, $\mathcal{K}_{\hat{W}}$ does not only average over the WWD outcomes $\ket{W_i}$, but also implicitly over the phase outcomes $\delta$.
If we average $\mathcal{K}_{\hat{W}}(\delta)$ over all phases with their respective probability, we reobtain $\mathcal{K}_{\hat{W}}$~\footnote{This can be seen by noting that $\int_0^{2\pi}\dd\delta p_i(\delta)P(\delta) = p_i$.}
\begin{align}
    \overline{\mathcal{K}_{\hat{W}}(\delta)} & \equiv \int_0^{2\pi}\mathcal{K}_{\hat{W}}(\delta)P(\delta)\dd\delta = \mathcal{K}_{\hat{W}} \, .
    \label{eq_knowledge_phase_average}
\end{align}
Considering the two previous examples of natural and canonical basis, their phase dependence can be calculated to
$\mathcal{K}_{\hat{E}}(\delta) = \sqrt{1-\mathcal{V}^2} = \mathcal{K}_{\hat{E}}$, and $\mathcal{K}_{\hat{N}}(\delta) = (1-\mathcal{V})/(1+\cos{(\delta)}\mathcal{V})$~\cite{Schilling:2012} also shown in Fig.~\ref{fig_all}~(d) and (e) for two different visibilities.
The canonical knowledge is independent of the phase $\delta$ of the QO and always equal to the optimal value permitted by the duality relation. This means that for the canonical observable the WW knowledge available is uncorrelated to the QO's position on the screen. 
On the other hand, for the natural observable $\hat{N}$ the knowledge and the QO's phase are correlated.
In particular, there is a region around the interference minimum 
where $\mathcal{K}_{\hat{N}}(\delta)$ exceeds the limit set by the duality relation, $\mathcal{K}_{\hat{N}}(\delta) > \sqrt{1-\mathcal{V}^2}$~\cite{Schilling:2012}.
Note, however, that this surpassing of the duality relation is only a local outcome, i.e., it appears for certain values of $\delta$.
Yet, on phase average, the duality relation is still fulfilled for the natural observable $\overline{\mathcal{K}_{\hat{N}}(\delta)}=\mathcal{K}_{\hat{N}}=1-\mathcal{V}\leq \sqrt{1-\mathcal{V}^2}$. The same holds true for any WWD observable if the basis remains fixed for all single experimental runs, see Eq.~\eqref{eq_knowledge_phase_average}.

\paragraph*{Feed-forward protocol.}\label{sec:feed-forward-protocol}

Motivated by the previous example,
we propose a feed-forward protocol aiming at increasing the WW knowledge beyond the duality relation limit by choosing a basis adapted for each experimental run. 
Within each run, the QO propagates through the double slit and interacts with the WWD; 
subsequently, the QO's phase $\delta$ is measured at the screen; 
we then choose - depending on the observed phase - the WWD observable $\hat{W}_\delta$ with eigenbasis $W_\delta = \{\ket{W_{\delta,i}},\, i=1,\ldots\}$ that maximizes $\mathcal{K}_{\hat{W}}(\delta)$ at the observed phase with respect to $\hat{W}$. 
Over multiple runs, this protocol returns the feed-forward (FF) which-way knowledge
\begin{align}
    \mathcal{K}_{FF}(\delta) \equiv \mathcal{K}_{\hat{W}_\delta}(\delta) = 2\sum_i p_{\delta,i}(\delta) q_{\delta,i}-1,
    \label{eq_knowledge_maximized}
\end{align}
where, in analogy to Eq. \eqref{eq_knowledge_phasedep}, we average over the possible WWD readout results $\{\ket{W_{\delta,i}},\, i=1,\ldots\}$
for each phase $\delta$ separately.
In contrast to Eq. \eqref{eq_knowledge_phasedep}, the basis set $W_\delta$ potentially differs for every value of $\delta$, what is indicated by the additional subscript $\delta$ in each quantity.

The difference between the proposed feed-forward protocol and a  fixed basis setup is depicted in Fig.~\ref{fig_all}(c) in a quantum circuit diagram.
Here, identifying the two pathways as the two states of a qubit, i.e., $\ket{\psi_{a}} \equiv \ket{0}$ and $\ket{\psi_{b}} \equiv \ket{1}$, 
the slit passage of Eq.~\eqref{eq_QO_wavefunc_slit} can be related to a Hadamard gate (H). 
The WWD subsequently stores partial information via controlled unitaries $U_\beta$ reflecting Eqs.~\eqref{eq_QO_WWD_entangledstate} and~\eqref{eq_WWD_std-basis} (e.g., via controlled rotations $R_y(2\theta)$ with $\beta = \sin\theta$); here, the bit flips (Pauli $X$ gates) are necessary to be sensitive for the state $\ket{\psi_{a}} \equiv \ket{0} $. 
Afterwards, the phase $\delta$ of the QO is measured (e.g., via a combination of a phase and a Hadamard gate simulating the accumulation of the phase and the combination of the two paths, respectively).
For a fixed basis $\hat{W}$ of the WWD, an optimization leads to the canonical knowledge $\mathcal{K}_{\hat{E}}$; 
however, via our protocol of feed-forwarding the measured phase $\delta$ of the QO, we can optimize the basis $\hat{W}_\delta$ at each phase leading to $\mathcal{K}_{FF}(\delta)$.

\paragraph*{Maximizing the which-way knowledge.}

In order to find the maximum feed-forward WW knowledge $\mathcal{K}_{FF}(\delta)$, we have to choose at each $\delta$ the observable $\hat{W}_\delta$ of all possible WWD observables that maximizes $\mathcal{K}_{\hat{W}}(\delta)$ with respect to $\hat{W}$.
To the best of our knowledge, there is no known analytic solution to this task.
Therefore, we will first consider a simplified version analytically and afterwards determine $\mathcal{K}_{FF}(\delta)$ in the non-simplified protocol numerically.

In a simplified version of the feed-forward protocol, 
we only choose between the natural and canonical observable such that the WW knowledge becomes maximized at the found phase, i.e., $\mathcal{K}_{\hat{N}/\hat{E}} (\delta) = \text{max} \left\{\mathcal{K}_{\hat{N}}(\delta), \mathcal{K}_{\hat{E}}(\delta) \right\}$.
From Fig.~\ref{fig_all}(d) and (e), we can infer that
away from the interference minimum at $\delta=\pi$, $\mathcal{K}_{\hat{N}/\hat{E}}(\delta)$ equals the canonical knowledge $\mathcal{K}_{\hat{E}}$, while close to the interference minimum, it equals the natural observable $\hat{N}$, whose knowledge exceeds the duality relation limit locally.
Consequently, the resulting knowledge becomes increased with respect to the canonical knowledge, i.e., $\mathcal{K}_{\hat{N}/\hat{E}}(\delta) \geq \mathcal{K}_{\hat{E}}$ for all phases $ \delta\in [0,2\pi]$.
For visibilities $0<\mathcal{V}<1$, 
this strictly transfers to the phase average $\overline{\mathcal{K}_{\hat{N}/\hat{E}}(\delta)}$ (calculated analytically in the end matter) such that it surpasses the canonical knowledge globally,
$\overline{\mathcal{K}_{\hat{N}/\hat{E}}(\delta)} > \mathcal{K}_{\hat{E}}$.

In the non-simplified feed-forward protocol, we determine the maximum feed-forward WW knowledge $\mathcal{K}_{FF}(\delta)$ numerically:
For fifty equally separated values of $\delta\in[0,2\pi]$, we calculate $\mathcal{K}_{\hat{W}}(\delta)$ for $50.000$ random orthogonal bases of the subspace $\text{span}\{\ket{00},\ket{10},\ket{01}\}$ \footnote{We can restrict ourselves to this subspace of $\mathcal{H}_\text{WWD}$ since due to the decomposition of Eq.~\eqref{eq_WWD_std-basis}, the detection probability of the state $\ket{11}$ would be zero, such that contributions along that vector do not influence the WW knowledge.} for a given value of $\mathcal{V}$.
In Fig.~\ref{fig_all}(d) and (e), we show the resulting maximum knowledge $\mathcal{K}_{FF}(\delta)$ for two different values of $\mathcal{V}$ as blue dots.
For most values of $\delta$, $\mathcal{K}_{FF}(\delta)$ follows the canonical one (close to the interference maxima) and natural one (close to the interference minima).
Only in a small interval between the two regions, there exists a different basis that increases the knowledge even further.

\paragraph{Surpassing of the duality relation.}

This knowledge increase (simplified and non-simplified protocol) does 
allow us to exceed the optimal value permitted by the duality relation \eqref{eq_duality} globally.
The excess provided by the feed-forward protocol can be quantified via computing $\overline{\mathcal{K}_{\ldots}(\delta)}^2 + \mathcal{V}^2$, as shown in Fig.~\ref{fig_all}(f) as a function of the visibility $\mathcal{V}\in[0,1]$.
Here, the canonical observable $\hat{E}$ sets the baseline (constant one) and a result greater than $1$ indicates a surpassing of the duality relation.
In the simplified protocol the maximum excess can be analytically calculated to be $\approx 1.016$ at $\mathcal{V}\approx 0.64$; while in the non-simplified protocol, the maximum excess numerically evaluates to $\approx 1.025$ at $\mathcal{V} \approx 0.59$~\footnote{This amount of surpassing is smaller than first expected from Fig.~\ref{fig_all}(d) and (e). We can understand this by noting that the largest WW knowledge increase is possible within the region around the interference minimum and thus weighted with the comparatively low probability of measuring the QO there [see Eq. \eqref{eq_knowledge_phase_average}].}.

The WW knowledges obtained via feed-forward of phase information $\mathcal{K}_{\hat{N}/\hat{E}}(\delta)$ and $\mathcal{K}_{FF}(\delta)$ thus exceed the duality relation limit of Eq.~\eqref{eq_duality} globally on phase average
\begin{align}
    \overline{\mathcal{K}_{\hat{N}/\hat{E}}(\delta)}^2 + \mathcal{V}^2 \geq 1 \quad\text{and}\quad \overline{\mathcal{K}_{FF}(\delta)}^2 + \mathcal{V}^2 \geq 1 \, ,
    \label{eq_duality_violation}
\end{align}
with the inequalities becoming strict for visibilities $0 < \mathcal{V} < 1$.

\paragraph*{Order of measurement.}

The design of our feed-forward protocol makes it clear that the order of measurement of the QO and WWD is crucial when we aim at
maximizing the WW knowledge.
Only in the case that the QO is measured first, we are able to 
strategically select the WWD observable
depending on the observed phase.
This allows us to optimize $\mathcal{K}_{\hat{W}}(\delta)$ for each phase individually, leading to the feed-forward WW knowledge $\mathcal{K}_{FF}(\delta)$.
In contrast, when reading out the WWD first, we fix the WWD observable before the phase of the QO is measured.
Here, our best candidate to maximize the WW knowledge is the canonical observable $\hat{E}$
leading to the canonical knowledge $\mathcal{K}_{\hat{E}}$ ~\cite{Englert:1996} (see Eq.~\eqref{eq_knowledge_canonical}).
Note, however, that the interference pattern $P(\delta)$ and thus the visibility and the wave-like behavior is independent of the measurement order (see end matter).

\paragraph*{Conclusion.}

In summary, we have seen that for a single run of Young's double-slit experiment, the resulting phase of the quantum object (QO) at the detection screen can be correlated to the which-way (WW) information found at the which-way detector (WWD).
Over multiple runs of the experiment, these correlations give rise to a phase-dependency of the WW knowledge, whose functional form depends on the WWD observable chosen. 
By feed-forwarding the phase information we can maximize the WW knowledge with respect to the WWD observable at each phase, leading to an increase of the WW knowledge beyond the duality relation limit at certain phases as well as on phase average. To achieve this, it is crucial to read out the QO before the WWD.

We stress that our central finding, i.e., the surpassing of the duality relation in Eq.~\eqref{eq_duality_violation}, is not in contradiction to the duality relation of Eq.~\eqref{eq_duality}~\cite{Englert:1996}. The latter is based on the assumption of choosing a fixed WWD observable over all runs of the experiment. By contrast, our feed-forward protocol allows for a new choice of the WWD observable for each single run of the experiment depending on the measured phase of the QO.
Our findings thus reveal how the outcome of a measurement in an appropriate feed-forward protocol can affect complementarity.

Future investigations may include an analytic solution for the maximum feed-forward WW knowledge $\mathcal{K}_{FF}(\delta)$, and a generalization of the feed-forward protocol to weighted interferometers as well as to multiple slits and/or mixed states.
It might also be interesting to investigate implications of our findings for potential applications in quantum computing, a field that frequently exploits such feed-forward processes.

\paragraph*{Acknowledgments.}

This work was funded by the Deutsche Forschungsgemeinschaft (DFG, German Research Foundation) – Project-ID 429529648 – TRR 306 QuCoLiMa (“Quantum Cooperativity of Light and Matter’’).
U.S., M.-O.P., and J.v.Z. acknowledge funding by the Erlangen Graduate School in Advanced Optical Technologies (SAOT) by the Bavarian State Ministry for Science and Art.

\bibliography{bibliography_PRL}

\begin{thebibliography}{29}%
\makeatletter
\providecommand \@ifxundefined [1]{%
 \@ifx{#1\undefined}
}%
\providecommand \@ifnum [1]{%
 \ifnum #1\expandafter \@firstoftwo
 \else \expandafter \@secondoftwo
 \fi
}%
\providecommand \@ifx [1]{%
 \ifx #1\expandafter \@firstoftwo
 \else \expandafter \@secondoftwo
 \fi
}%
\providecommand \natexlab [1]{#1}%
\providecommand \enquote  [1]{``#1''}%
\providecommand \bibnamefont  [1]{#1}%
\providecommand \bibfnamefont [1]{#1}%
\providecommand \citenamefont [1]{#1}%
\providecommand \href@noop [0]{\@secondoftwo}%
\providecommand \href [0]{\begingroup \@sanitize@url \@href}%
\providecommand \@href[1]{\@@startlink{#1}\@@href}%
\providecommand \@@href[1]{\endgroup#1\@@endlink}%
\providecommand \@sanitize@url [0]{\catcode `\\12\catcode `\$12\catcode `\&12\catcode `\#12\catcode `\^12\catcode `\_12\catcode `\%12\relax}%
\providecommand \@@startlink[1]{}%
\providecommand \@@endlink[0]{}%
\providecommand \url  [0]{\begingroup\@sanitize@url \@url }%
\providecommand \@url [1]{\endgroup\@href {#1}{\urlprefix }}%
\providecommand \urlprefix  [0]{URL }%
\providecommand \Eprint [0]{\href }%
\providecommand \doibase [0]{https://doi.org/}%
\providecommand \selectlanguage [0]{\@gobble}%
\providecommand \bibinfo  [0]{\@secondoftwo}%
\providecommand \bibfield  [0]{\@secondoftwo}%
\providecommand \translation [1]{[#1]}%
\providecommand \BibitemOpen [0]{}%
\providecommand \bibitemStop [0]{}%
\providecommand \bibitemNoStop [0]{.\EOS\space}%
\providecommand \EOS [0]{\spacefactor3000\relax}%
\providecommand \BibitemShut  [1]{\csname bibitem#1\endcsname}%
\let\auto@bib@innerbib\@empty
\bibitem [{\citenamefont {Bohr}(1949)}]{Bohr:1949}%
  \BibitemOpen
  \bibfield  {author} {\bibinfo {author} {\bibfnamefont {N.}~\bibnamefont {Bohr}},\ }\bibinfo {title} {Discussion with {E}instein on epistemological problems in atomic physics}\ (\bibinfo  {publisher} {Evanston},\ \bibinfo {year} {1949})\ pp.\ \bibinfo {pages} {200--241}\BibitemShut {NoStop}%
\bibitem [{\citenamefont {Englert}(1996)}]{Englert:1996}%
  \BibitemOpen
  \bibfield  {author} {\bibinfo {author} {\bibfnamefont {B.-G.}\ \bibnamefont {Englert}},\ }\bibfield  {title} {\bibinfo {title} {Fringe visibility and which-way information: An inequality},\ }\href@noop {} {\bibfield  {journal} {\bibinfo  {journal} {Phys. Rev. Lett.}\ }\textbf {\bibinfo {volume} {77}},\ \bibinfo {pages} {2154} (\bibinfo {year} {1996})}\BibitemShut {NoStop}%
\bibitem [{\citenamefont {Jaeger}\ \emph {et~al.}(1995)\citenamefont {Jaeger}, \citenamefont {Shimony},\ and\ \citenamefont {Vaidman}}]{Jaeger:1995}%
  \BibitemOpen
  \bibfield  {author} {\bibinfo {author} {\bibfnamefont {G.}~\bibnamefont {Jaeger}}, \bibinfo {author} {\bibfnamefont {A.}~\bibnamefont {Shimony}},\ and\ \bibinfo {author} {\bibfnamefont {L.}~\bibnamefont {Vaidman}},\ }\bibfield  {title} {\bibinfo {title} {Two interferometric complementarities},\ }\href@noop {} {\bibfield  {journal} {\bibinfo  {journal} {Phys. Rev. A}\ }\textbf {\bibinfo {volume} {51}},\ \bibinfo {pages} {54} (\bibinfo {year} {1995})}\BibitemShut {NoStop}%
\bibitem [{\citenamefont {D{\"u}rr}\ \emph {et~al.}(1998)\citenamefont {D{\"u}rr}, \citenamefont {Nonn},\ and\ \citenamefont {Rempe}}]{Durr:1998}%
  \BibitemOpen
  \bibfield  {author} {\bibinfo {author} {\bibfnamefont {S.}~\bibnamefont {D{\"u}rr}}, \bibinfo {author} {\bibfnamefont {T.}~\bibnamefont {Nonn}},\ and\ \bibinfo {author} {\bibfnamefont {G.}~\bibnamefont {Rempe}},\ }\bibfield  {title} {\bibinfo {title} {Origin of quantum-mechanical complementarity probed by a `which-way' experiment in an atom interferometer},\ }\href@noop {} {\bibfield  {journal} {\bibinfo  {journal} {Nature}\ }\textbf {\bibinfo {volume} {395}},\ \bibinfo {pages} {33} (\bibinfo {year} {1998})}\BibitemShut {NoStop}%
\bibitem [{\citenamefont {D\"urr}\ \emph {et~al.}(1998)\citenamefont {D\"urr}, \citenamefont {Nonn},\ and\ \citenamefont {Rempe}}]{Durr:1998a}%
  \BibitemOpen
  \bibfield  {author} {\bibinfo {author} {\bibfnamefont {S.}~\bibnamefont {D\"urr}}, \bibinfo {author} {\bibfnamefont {T.}~\bibnamefont {Nonn}},\ and\ \bibinfo {author} {\bibfnamefont {G.}~\bibnamefont {Rempe}},\ }\bibfield  {title} {\bibinfo {title} {Fringe visibility and which-way information in an atom interferometer},\ }\href@noop {} {\bibfield  {journal} {\bibinfo  {journal} {Phys. Rev. Lett.}\ }\textbf {\bibinfo {volume} {81}},\ \bibinfo {pages} {5705} (\bibinfo {year} {1998})}\BibitemShut {NoStop}%
\bibitem [{\citenamefont {Buks}\ \emph {et~al.}(1998)\citenamefont {Buks}, \citenamefont {Schuster}, \citenamefont {Heiblum}, \citenamefont {Mahalu},\ and\ \citenamefont {Umansky}}]{Buks:1998}%
  \BibitemOpen
  \bibfield  {author} {\bibinfo {author} {\bibfnamefont {E.}~\bibnamefont {Buks}}, \bibinfo {author} {\bibfnamefont {R.}~\bibnamefont {Schuster}}, \bibinfo {author} {\bibfnamefont {M.}~\bibnamefont {Heiblum}}, \bibinfo {author} {\bibfnamefont {D.}~\bibnamefont {Mahalu}},\ and\ \bibinfo {author} {\bibfnamefont {V.}~\bibnamefont {Umansky}},\ }\bibfield  {title} {\bibinfo {title} {Dephasing in electron interference by a `which-path' detector},\ }\href@noop {} {\bibfield  {journal} {\bibinfo  {journal} {Nature}\ }\textbf {\bibinfo {volume} {391}},\ \bibinfo {pages} {871} (\bibinfo {year} {1998})}\BibitemShut {NoStop}%
\bibitem [{\citenamefont {Schwindt}\ \emph {et~al.}(1999)\citenamefont {Schwindt}, \citenamefont {Kwiat},\ and\ \citenamefont {Englert}}]{Schwindt:1999}%
  \BibitemOpen
  \bibfield  {author} {\bibinfo {author} {\bibfnamefont {P.~D.~D.}\ \bibnamefont {Schwindt}}, \bibinfo {author} {\bibfnamefont {P.~G.}\ \bibnamefont {Kwiat}},\ and\ \bibinfo {author} {\bibfnamefont {B.-G.}\ \bibnamefont {Englert}},\ }\bibfield  {title} {\bibinfo {title} {Quantitative wave-particle duality and nonerasing quantum erasure},\ }\href@noop {} {\bibfield  {journal} {\bibinfo  {journal} {Phys. Rev. A}\ }\textbf {\bibinfo {volume} {60}},\ \bibinfo {pages} {4285} (\bibinfo {year} {1999})}\BibitemShut {NoStop}%
\bibitem [{\citenamefont {Pryde}\ \emph {et~al.}(2004)\citenamefont {Pryde}, \citenamefont {O'Brien}, \citenamefont {White}, \citenamefont {Bartlett},\ and\ \citenamefont {Ralph}}]{Pryde:2004}%
  \BibitemOpen
  \bibfield  {author} {\bibinfo {author} {\bibfnamefont {G.~J.}\ \bibnamefont {Pryde}}, \bibinfo {author} {\bibfnamefont {J.~L.}\ \bibnamefont {O'Brien}}, \bibinfo {author} {\bibfnamefont {A.~G.}\ \bibnamefont {White}}, \bibinfo {author} {\bibfnamefont {S.~D.}\ \bibnamefont {Bartlett}},\ and\ \bibinfo {author} {\bibfnamefont {T.~C.}\ \bibnamefont {Ralph}},\ }\bibfield  {title} {\bibinfo {title} {Measuring a photonic qubit without destroying it},\ }\href@noop {} {\bibfield  {journal} {\bibinfo  {journal} {Phys. Rev. Lett.}\ }\textbf {\bibinfo {volume} {92}},\ \bibinfo {pages} {190402} (\bibinfo {year} {2004})}\BibitemShut {NoStop}%
\bibitem [{\citenamefont {Peng}\ \emph {et~al.}(2005)\citenamefont {Peng}, \citenamefont {Zhu}, \citenamefont {Suter}, \citenamefont {Du}, \citenamefont {Liu},\ and\ \citenamefont {Gao}}]{Peng:2005}%
  \BibitemOpen
  \bibfield  {author} {\bibinfo {author} {\bibfnamefont {X.}~\bibnamefont {Peng}}, \bibinfo {author} {\bibfnamefont {X.}~\bibnamefont {Zhu}}, \bibinfo {author} {\bibfnamefont {D.}~\bibnamefont {Suter}}, \bibinfo {author} {\bibfnamefont {J.}~\bibnamefont {Du}}, \bibinfo {author} {\bibfnamefont {M.}~\bibnamefont {Liu}},\ and\ \bibinfo {author} {\bibfnamefont {K.}~\bibnamefont {Gao}},\ }\bibfield  {title} {\bibinfo {title} {Quantification of complementarity in multiqubit systems},\ }\href@noop {} {\bibfield  {journal} {\bibinfo  {journal} {Phys. Rev. A}\ }\textbf {\bibinfo {volume} {72}},\ \bibinfo {pages} {052109} (\bibinfo {year} {2005})}\BibitemShut {NoStop}%
\bibitem [{\citenamefont {Jacques}\ \emph {et~al.}(2008)\citenamefont {Jacques}, \citenamefont {Wu}, \citenamefont {Grosshans}, \citenamefont {Treussart}, \citenamefont {Grangier}, \citenamefont {Aspect},\ and\ \citenamefont {Roch}}]{Jacques:2008}%
  \BibitemOpen
  \bibfield  {author} {\bibinfo {author} {\bibfnamefont {V.}~\bibnamefont {Jacques}}, \bibinfo {author} {\bibfnamefont {E.}~\bibnamefont {Wu}}, \bibinfo {author} {\bibfnamefont {F.}~\bibnamefont {Grosshans}}, \bibinfo {author} {\bibfnamefont {F.}~\bibnamefont {Treussart}}, \bibinfo {author} {\bibfnamefont {P.}~\bibnamefont {Grangier}}, \bibinfo {author} {\bibfnamefont {A.}~\bibnamefont {Aspect}},\ and\ \bibinfo {author} {\bibfnamefont {J.-F.}\ \bibnamefont {Roch}},\ }\bibfield  {title} {\bibinfo {title} {Delayed-choice test of quantum complementarity with interfering single photons},\ }\href@noop {} {\bibfield  {journal} {\bibinfo  {journal} {Phys. Rev. Lett.}\ }\textbf {\bibinfo {volume} {100}},\ \bibinfo {pages} {220402} (\bibinfo {year} {2008})}\BibitemShut {NoStop}%
\bibitem [{\citenamefont {Barbieri}\ \emph {et~al.}(2009)\citenamefont {Barbieri}, \citenamefont {Goggin}, \citenamefont {Almeida}, \citenamefont {Lanyon},\ and\ \citenamefont {White}}]{Barbieri:2009}%
  \BibitemOpen
  \bibfield  {author} {\bibinfo {author} {\bibfnamefont {M.}~\bibnamefont {Barbieri}}, \bibinfo {author} {\bibfnamefont {M.~E.}\ \bibnamefont {Goggin}}, \bibinfo {author} {\bibfnamefont {M.~P.}\ \bibnamefont {Almeida}}, \bibinfo {author} {\bibfnamefont {B.~P.}\ \bibnamefont {Lanyon}},\ and\ \bibinfo {author} {\bibfnamefont {A.~G.}\ \bibnamefont {White}},\ }\bibfield  {title} {\bibinfo {title} {Complementarity in variable strength quantum non-demolition measurements},\ }\href@noop {} {\bibfield  {journal} {\bibinfo  {journal} {New Journal of Physics}\ }\textbf {\bibinfo {volume} {11}},\ \bibinfo {pages} {093012} (\bibinfo {year} {2009})}\BibitemShut {NoStop}%
\bibitem [{\citenamefont {Bj\"ork}\ and\ \citenamefont {Karlsson}(1998)}]{Bjork:1998}%
  \BibitemOpen
  \bibfield  {author} {\bibinfo {author} {\bibfnamefont {G.}~\bibnamefont {Bj\"ork}}\ and\ \bibinfo {author} {\bibfnamefont {A.}~\bibnamefont {Karlsson}},\ }\bibfield  {title} {\bibinfo {title} {Complementarity and quantum erasure in welcher weg experiments},\ }\href@noop {} {\bibfield  {journal} {\bibinfo  {journal} {Phys. Rev. A}\ }\textbf {\bibinfo {volume} {58}},\ \bibinfo {pages} {3477} (\bibinfo {year} {1998})}\BibitemShut {NoStop}%
\bibitem [{\citenamefont {Abranyos}\ \emph {et~al.}(1999)\citenamefont {Abranyos}, \citenamefont {Jakob},\ and\ \citenamefont {Bergou}}]{Abranyos:1999}%
  \BibitemOpen
  \bibfield  {author} {\bibinfo {author} {\bibfnamefont {Y.}~\bibnamefont {Abranyos}}, \bibinfo {author} {\bibfnamefont {M.}~\bibnamefont {Jakob}},\ and\ \bibinfo {author} {\bibfnamefont {J.}~\bibnamefont {Bergou}},\ }\bibfield  {title} {\bibinfo {title} {Interference and partial which-way information: A quantitative test of duality in two-atom resonance},\ }\href@noop {} {\bibfield  {journal} {\bibinfo  {journal} {Phys. Rev. A}\ }\textbf {\bibinfo {volume} {61}},\ \bibinfo {pages} {013804} (\bibinfo {year} {1999})}\BibitemShut {NoStop}%
\bibitem [{\citenamefont {Englert}\ and\ \citenamefont {Bergou}(2000)}]{Englert:2000}%
  \BibitemOpen
  \bibfield  {author} {\bibinfo {author} {\bibfnamefont {B.-G.}\ \bibnamefont {Englert}}\ and\ \bibinfo {author} {\bibfnamefont {J.~A.}\ \bibnamefont {Bergou}},\ }\bibfield  {title} {\bibinfo {title} {Quantitative quantum erasure1dedicated to marlan scully on the occasion of his 60th birthday.1},\ }\href@noop {} {\bibfield  {journal} {\bibinfo  {journal} {Optics Communications}\ }\textbf {\bibinfo {volume} {179}},\ \bibinfo {pages} {337} (\bibinfo {year} {2000})}\BibitemShut {NoStop}%
\bibitem [{\citenamefont {Miniatura}\ \emph {et~al.}(2007)\citenamefont {Miniatura}, \citenamefont {M\"uller}, \citenamefont {Lu}, \citenamefont {Wang},\ and\ \citenamefont {Englert}}]{Miniatura:2007}%
  \BibitemOpen
  \bibfield  {author} {\bibinfo {author} {\bibfnamefont {C.}~\bibnamefont {Miniatura}}, \bibinfo {author} {\bibfnamefont {C.~A.}\ \bibnamefont {M\"uller}}, \bibinfo {author} {\bibfnamefont {Y.}~\bibnamefont {Lu}}, \bibinfo {author} {\bibfnamefont {G.}~\bibnamefont {Wang}},\ and\ \bibinfo {author} {\bibfnamefont {B.-G.}\ \bibnamefont {Englert}},\ }\bibfield  {title} {\bibinfo {title} {Path distinguishability in double scattering of light by atoms},\ }\href@noop {} {\bibfield  {journal} {\bibinfo  {journal} {Phys. Rev. A}\ }\textbf {\bibinfo {volume} {76}},\ \bibinfo {pages} {022101} (\bibinfo {year} {2007})}\BibitemShut {NoStop}%
\bibitem [{\citenamefont {Erez}\ \emph {et~al.}(2009)\citenamefont {Erez}, \citenamefont {Jacobs},\ and\ \citenamefont {Kurizki}}]{Erez:2009}%
  \BibitemOpen
  \bibfield  {author} {\bibinfo {author} {\bibfnamefont {N.}~\bibnamefont {Erez}}, \bibinfo {author} {\bibfnamefont {D.}~\bibnamefont {Jacobs}},\ and\ \bibinfo {author} {\bibfnamefont {G.}~\bibnamefont {Kurizki}},\ }\bibfield  {title} {\bibinfo {title} {Operational path--phase complementarity in single-photon interferometry},\ }\href@noop {} {\bibfield  {journal} {\bibinfo  {journal} {Journal of Physics B: Atomic, Molecular and Optical Physics}\ }\textbf {\bibinfo {volume} {42}},\ \bibinfo {pages} {114006} (\bibinfo {year} {2009})}\BibitemShut {NoStop}%
\bibitem [{\citenamefont {Bolduc}\ \emph {et~al.}(2014)\citenamefont {Bolduc}, \citenamefont {Leach}, \citenamefont {Miatto}, \citenamefont {Leuchs},\ and\ \citenamefont {Boyd}}]{Bolduc:2014}%
  \BibitemOpen
  \bibfield  {author} {\bibinfo {author} {\bibfnamefont {E.}~\bibnamefont {Bolduc}}, \bibinfo {author} {\bibfnamefont {J.}~\bibnamefont {Leach}}, \bibinfo {author} {\bibfnamefont {F.~M.}\ \bibnamefont {Miatto}}, \bibinfo {author} {\bibfnamefont {G.}~\bibnamefont {Leuchs}},\ and\ \bibinfo {author} {\bibfnamefont {R.~W.}\ \bibnamefont {Boyd}},\ }\bibfield  {title} {\bibinfo {title} {Fair sampling perspective on an apparent violation of duality},\ }\href@noop {} {\bibfield  {journal} {\bibinfo  {journal} {Proceedings of the National Academy of Sciences}\ }\textbf {\bibinfo {volume} {111}},\ \bibinfo {pages} {12337} (\bibinfo {year} {2014})}\BibitemShut {NoStop}%
\bibitem [{\citenamefont {Leach}\ \emph {et~al.}(2016)\citenamefont {Leach}, \citenamefont {Bolduc}, \citenamefont {Miatto}, \citenamefont {Pich{\'e}}, \citenamefont {Leuchs},\ and\ \citenamefont {Boyd}}]{Leach:2016}%
  \BibitemOpen
  \bibfield  {author} {\bibinfo {author} {\bibfnamefont {J.}~\bibnamefont {Leach}}, \bibinfo {author} {\bibfnamefont {E.}~\bibnamefont {Bolduc}}, \bibinfo {author} {\bibfnamefont {F.~M.}\ \bibnamefont {Miatto}}, \bibinfo {author} {\bibfnamefont {K.}~\bibnamefont {Pich{\'e}}}, \bibinfo {author} {\bibfnamefont {G.}~\bibnamefont {Leuchs}},\ and\ \bibinfo {author} {\bibfnamefont {R.~W.}\ \bibnamefont {Boyd}},\ }\bibfield  {title} {\bibinfo {title} {The duality principle in the presence of postselection},\ }\href@noop {} {\bibfield  {journal} {\bibinfo  {journal} {Scientific Reports}\ }\textbf {\bibinfo {volume} {6}},\ \bibinfo {pages} {19944} (\bibinfo {year} {2016})}\BibitemShut {NoStop}%
\bibitem [{\citenamefont {Chen}\ \emph {et~al.}(2022)\citenamefont {Chen}, \citenamefont {Zhang}, \citenamefont {Zhao}, \citenamefont {Wu}, \citenamefont {Fang}, \citenamefont {Yang},\ and\ \citenamefont {Nori}}]{Chen:2022}%
  \BibitemOpen
  \bibfield  {author} {\bibinfo {author} {\bibfnamefont {D.-X.}\ \bibnamefont {Chen}}, \bibinfo {author} {\bibfnamefont {Y.}~\bibnamefont {Zhang}}, \bibinfo {author} {\bibfnamefont {J.-L.}\ \bibnamefont {Zhao}}, \bibinfo {author} {\bibfnamefont {Q.-C.}\ \bibnamefont {Wu}}, \bibinfo {author} {\bibfnamefont {Y.-L.}\ \bibnamefont {Fang}}, \bibinfo {author} {\bibfnamefont {C.-P.}\ \bibnamefont {Yang}},\ and\ \bibinfo {author} {\bibfnamefont {F.}~\bibnamefont {Nori}},\ }\bibfield  {title} {\bibinfo {title} {Experimental investigation of wave-particle duality relations in asymmetric beam interference},\ }\href@noop {} {\bibfield  {journal} {\bibinfo  {journal} {npj Quantum Information}\ }\textbf {\bibinfo {volume} {8}},\ \bibinfo {pages} {101} (\bibinfo {year} {2022})}\BibitemShut {NoStop}%
\bibitem [{\citenamefont {Schilling}\ and\ \citenamefont {{von Zanthier}}(2012)}]{Schilling:2012}%
  \BibitemOpen
  \bibfield  {author} {\bibinfo {author} {\bibfnamefont {U.}~\bibnamefont {Schilling}}\ and\ \bibinfo {author} {\bibfnamefont {J.}~\bibnamefont {{von Zanthier}}},\ }\bibfield  {title} {\bibinfo {title} {Phase-dependent which-way information},\ }\href@noop {} {\bibfield  {journal} {\bibinfo  {journal} {Physics Letters A}\ }\textbf {\bibinfo {volume} {376}},\ \bibinfo {pages} {3479} (\bibinfo {year} {2012})}\BibitemShut {NoStop}%
\bibitem [{\citenamefont {Bera}\ \emph {et~al.}(2015)\citenamefont {Bera}, \citenamefont {Qureshi}, \citenamefont {Siddiqui},\ and\ \citenamefont {Pati}}]{Bera:2015}%
  \BibitemOpen
  \bibfield  {author} {\bibinfo {author} {\bibfnamefont {M.~N.}\ \bibnamefont {Bera}}, \bibinfo {author} {\bibfnamefont {T.}~\bibnamefont {Qureshi}}, \bibinfo {author} {\bibfnamefont {M.~A.}\ \bibnamefont {Siddiqui}},\ and\ \bibinfo {author} {\bibfnamefont {A.~K.}\ \bibnamefont {Pati}},\ }\bibfield  {title} {\bibinfo {title} {Duality of quantum coherence and path distinguishability},\ }\href@noop {} {\bibfield  {journal} {\bibinfo  {journal} {Phys. Rev. A}\ }\textbf {\bibinfo {volume} {92}},\ \bibinfo {pages} {012118} (\bibinfo {year} {2015})}\BibitemShut {NoStop}%
\bibitem [{\citenamefont {Spegel-Lexne}\ \emph {et~al.}(2024)\citenamefont {Spegel-Lexne}, \citenamefont {G{\'o}mez}, \citenamefont {Argillander}, \citenamefont {Paw{\l}owski}, \citenamefont {Dieguez}, \citenamefont {Alarc{\'o}n},\ and\ \citenamefont {Xavier}}]{Spegel-Lexne:2024}%
  \BibitemOpen
  \bibfield  {author} {\bibinfo {author} {\bibfnamefont {D.}~\bibnamefont {Spegel-Lexne}}, \bibinfo {author} {\bibfnamefont {S.}~\bibnamefont {G{\'o}mez}}, \bibinfo {author} {\bibfnamefont {J.}~\bibnamefont {Argillander}}, \bibinfo {author} {\bibfnamefont {M.}~\bibnamefont {Paw{\l}owski}}, \bibinfo {author} {\bibfnamefont {P.~R.}\ \bibnamefont {Dieguez}}, \bibinfo {author} {\bibfnamefont {A.}~\bibnamefont {Alarc{\'o}n}},\ and\ \bibinfo {author} {\bibfnamefont {G.~B.}\ \bibnamefont {Xavier}},\ }\bibfield  {title} {\bibinfo {title} {Experimental demonstration of the equivalence of entropic uncertainty with wave-particle duality},\ }\href@noop {} {\bibfield  {journal} {\bibinfo  {journal} {Science Advances}\ }\textbf {\bibinfo {volume} {10}},\ \bibinfo {pages} {eadr2007} (\bibinfo {year} {2024})}\BibitemShut {NoStop}%
\bibitem [{\citenamefont {{Jiang}}\ \emph {et~al.}(2025)\citenamefont {{Jiang}}, \citenamefont {{Hochrainer}}, \citenamefont {{Kysela}}, \citenamefont {{Erhard}}, \citenamefont {{Gu}}, \citenamefont {{Yu}},\ and\ \citenamefont {{Zeilinger}}}]{Jiang:2025}%
  \BibitemOpen
  \bibfield  {author} {\bibinfo {author} {\bibfnamefont {X.}~\bibnamefont {{Jiang}}}, \bibinfo {author} {\bibfnamefont {A.}~\bibnamefont {{Hochrainer}}}, \bibinfo {author} {\bibfnamefont {J.}~\bibnamefont {{Kysela}}}, \bibinfo {author} {\bibfnamefont {M.}~\bibnamefont {{Erhard}}}, \bibinfo {author} {\bibfnamefont {X.}~\bibnamefont {{Gu}}}, \bibinfo {author} {\bibfnamefont {Y.}~\bibnamefont {{Yu}}},\ and\ \bibinfo {author} {\bibfnamefont {A.}~\bibnamefont {{Zeilinger}}},\ }\bibfield  {title} {\bibinfo {title} {{Subjective nature of path information in quantum mechanics}},\ }\href@noop {} {\bibfield  {journal} {\bibinfo  {journal} {arXiv e-prints}\ ,\ \bibinfo {pages} {arXiv:2505.05930}} (\bibinfo {year} {2025})}\BibitemShut {NoStop}%
\bibitem [{\citenamefont {Scully}\ \emph {et~al.}(1991)\citenamefont {Scully}, \citenamefont {Englert},\ and\ \citenamefont {Walther}}]{Scully:1991}%
  \BibitemOpen
  \bibfield  {author} {\bibinfo {author} {\bibfnamefont {M.~O.}\ \bibnamefont {Scully}}, \bibinfo {author} {\bibfnamefont {B.-G.}\ \bibnamefont {Englert}},\ and\ \bibinfo {author} {\bibfnamefont {H.}~\bibnamefont {Walther}},\ }\bibfield  {title} {\bibinfo {title} {Quantum optical tests of complementarity},\ }\href@noop {} {\bibfield  {journal} {\bibinfo  {journal} {Nature}\ }\textbf {\bibinfo {volume} {351}},\ \bibinfo {pages} {111} (\bibinfo {year} {1991})}\BibitemShut {NoStop}%
\bibitem [{Note1()}]{Note1}%
  \BibitemOpen
  \bibinfo {note} {In general, $\mathinner {|{\chi _{a,b}}\rangle }$ may have a relative phase $e^{i\phi }$, which besides shifting the interference pattern by a constant amount $\phi $ is without any further effect.}\BibitemShut {Stop}%
\bibitem [{Note2()}]{Note2}%
  \BibitemOpen
  \bibinfo {note} {Strictly speaking, the WWD Hilbert space has dimension $\protect \text {dim}(\protect \mathbb {C}^2\otimes \protect \mathbb {C}^2) = 4$ and $\{\mathinner {|{\chi _a}\rangle },\mathinner {|{\chi _b}\rangle }\}$ alone cannot form a basis. However, in case of orthogonal $\mathinner {|{\chi _{a/b}}\rangle }$, this set can be completed to an eigenbasis of a WWD observable that returns full WW information.}\BibitemShut {Stop}%
\bibitem [{Note3()}]{Note3}%
  \BibitemOpen
  \bibinfo {note} {This can be seen by noting that $\DOTSI \intop \ilimits@ _0^{2\pi }\protect \mathrm {d}\delta p_i(\delta )P(\delta ) = p_i$.}\BibitemShut {Stop}%
\bibitem [{Note4()}]{Note4}%
  \BibitemOpen
  \bibinfo {note} {We can restrict ourselves to this subspace of $\protect \mathcal {H}_\protect \text {WWD}$ since due to the decomposition of Eq.~\protect \eqref {eq_WWD_std-basis}, the detection probability of the state $\mathinner {|{11}\rangle }$ would be zero, such that contributions along that vector do not influence the WW knowledge.}\BibitemShut {Stop}%
\bibitem [{Note5()}]{Note5}%
  \BibitemOpen
  \bibinfo {note} {This amount of surpassing is smaller than first expected from Fig.~\ref {fig_all}(d) and (e). We can understand this by noting that the largest WW knowledge increase is possible within the region around the interference minimum and thus weighted with the comparatively low probability of measuring the QO there [see Eq. \protect \eqref {eq_knowledge_phase_average}].}\BibitemShut {Stop}%
\end{thebibliography}%

\onecolumngrid
\section*{END MATTER}
\twocolumngrid

\paragraph*{Analytic expressions for the simplified feed-forward protocol.}

The crossings of natural and canonical knowledge occur at $\delta_*$ and $2\pi-\delta_*$ with 
$$\cos(\delta_*)= (\sqrt{(1-\mathcal{V})/(1+\mathcal{V})} - 1)/\mathcal{V} \, $$ 
such that
\begin{align*}
\text{for} \, [0,\delta_*]: \quad &\mathcal{K}_{\hat{E}} \geq\mathcal{K}_{\hat{N}}\, , \\
\text{for} \, [\delta_*,2\pi-\delta_*]: \quad &\mathcal{K}_{\hat{N}} \geq\mathcal{K}_{\hat{E}}\, , \\
\text{for} \, [2\pi-\delta_*,2\pi]: \quad &\mathcal{K}_{\hat{E}} \geq\mathcal{K}_{\hat{N}} \, .
\end{align*}
Splitting the integral of the phase average accordingly into the three regions above, we can calculate the maximized knowledge in the simplified version of the feed forward protocol to
\begin{align*}
\overline{\mathcal{K}_{\hat{N}/\hat{E}}(\delta)}  = \frac{1}{2\pi} \Big[ &2 \sqrt{1-\mathcal{V}^2} (\delta_* + \mathcal{V} \sin\delta_*)\\
&  + 2 (1-\mathcal{V}) (\pi - \delta_*) \Big] ,
\end{align*}
which is strictly greater than the limit set by the duality relation ($\sqrt{1-\mathcal{V}^2}$) for visibilities $0<\mathcal{V}<1$.

\paragraph*{Measurement order and interference pattern.}

In case the QO is measured first, the interference pattern $P(\delta)$ is given by Eq. \eqref{eq_QO_interference_WWD}. 
In case the WWD is measured first, the entangled state $\ket{\Psi^{(f)}}$ is reduced to $\ket{\psi_i}$, and the interfering QO follows the probability distribution given by
$|\langle\psi_\delta|\psi_i\rangle|^2$.
However, to obtain the interference pattern, we have to perform multiple runs of the experiment and thus average over the possible WWD outcomes $\ket{W_i}$, returning the original interference pattern $\sum_i|\langle\psi_\delta|\psi_i\rangle|^2 = P(\delta)$ of Eq.~\eqref{eq_QO_interference_WWD}.
The measurement order thus does not alter the interference pattern $P(\delta)$. 

\paragraph*{Experimental verifiability.}

For the experimental feasibility of the results, the implementation of the phase-dependent WW knowledge operator $\mathcal{K}_{\hat{W}}(\delta)$ 
is particularly crucial.
In a Young interferometer with the QO recorded first on the screen, the correlation $\mathcal{K}_{\hat{W}}(\delta)$ can no longer be verified using a setup as shown in Fig.~\ref{fig_all}(b), as the QO has already been registered at the screen.
That restriction can be circumvented by realizing that the same formalism to derive $\mathcal{K}_{\hat{W}}(\delta)$ can be used in a Mach-Zehnder interferometer (MZI).
There, the phase $\delta$ is fixed by the setup and can be tuned by a phase shifter in one arm of the MZI, which sits before the WWD.
Hence, the QO accumulates the phase $\delta$ \textit{before} the interaction with the WWD and without being registered at the screen.
Thus, the WWD readout result is already correlated to the phase $\delta$ and the resulting correlation $\mathcal{K}_{\hat{W}}(\delta)$ can be checked by subsequently registering the QO within the interferometer arms $a$ and $b$ of the MZI.

\end{document}